\documentclass[12pt,a4paper]{elsarticle}
\usepackage[margin=2.5cm]{geometry}
\usepackage[T1]{fontenc}
\usepackage{lmodern}
\usepackage{microtype,url}
\usepackage[dvipsnames]{xcolor}
\usepackage{amsfonts,amsmath,amssymb,bm,mathtools}
\DeclareSymbolFont{vectors}{OML}{lmm}{b}{it}
\DeclareSymbolFont{tensors}{OT1}{lmss}{bx}{it}
\DeclareSymbolFontAlphabet{\tens}{tensors}
\DeclareSymbolFontAlphabet{\mathvec}{vectors} % Vectors
\usepackage[colorlinks=true,allcolors=blue]{hyperref}
\usepackage{siunitx}
\usepackage[noabbrev]{cleveref}

\crefname{section}{}{}
\crefname{equation}{}{}
\crefname{figure}{}{}
\crefname{table}{}{}
\crefname{appendix}{}{}
\crefname{chapter}{}{}
\numberwithin{equation}{section}
\numberwithin{figure}{section}
\numberwithin{table}{section}
\usepackage[section]{placeins}
\usepackage[mathlines]{lineno}
\usepackage[onehalfspacing]{setspace}

\usepackage{graphicx,subcaption}
\usepackage{tikz,pgfplots}
\pgfplotsset{compat=1.16}
\usepgfplotslibrary{statistics}
\usepackage{booktabs}

\DeclareRobustCommand{\d}{\relax\ifmmode\mathrm{d}\else\expandafter\@@d\fi}
\DeclareRobustCommand{\D}{\relax\ifmmode\mathrm{D}\else\expandafter\@@d\fi}
\DeclareRobustCommand{\e}{\relax\ifmmode\mathrm{e}\else\error\fi}
\DeclareRobustCommand{\i}{\relax\ifmmode\mathrm{i}\else\expandafter\@@i\fi}
\DeclareRobustCommand{\uppartial}{\text{\rotatebox[origin=t]{20}{\scalebox{0.95}[1]{\(\partial\)}}}\hspace{-1pt}}

\newcommand{\fd}[2]{\frac{\d #1}{\d #2}}

\newcommand{\pd}[2]{\frac{\uppartial #1}{\uppartial #2}}

\newcommand{\pdn}[3]{\frac{\uppartial^#3 #1}{\uppartial #2^#3}}

\newcommand{\df}[1]{\,\d#1}

\DeclarePairedDelimiterX{\abs}[1]{\lvert}{\rvert}{#1}
\DeclarePairedDelimiterX{\norm}[1]{\lVert}{\rVert}{#1}

\biboptions{sort&compress}

\begin{document}
%%%%%%%%%%%%%%%%%%%%%%%%
%%%%% Front Matter %%%%%
%%%%%%%%%%%%%%%%%%%%%%%%
\begin{frontmatter}
\title{The Effect of Geometry on Survival and Extinction in a Moving-Boundary Problem Motivated by the Fisher--KPP Equation}
% \author[qut]{Alexander K. Y. Tam\thanks{Corresponding author: \href{mailto:alexander.tam@qut.edu.au}{\texttt{alexander.tam@qut.edu.au}}}}
\author[qut]{Alexander K. Y. Tam}
\author[qut]{Matthew J. Simpson}
\address[qut]{School of Mathematical Sciences, Queensland University of Technology, Brisbane QLD 4000, Australia.}
%%%%%%%%%%%%%%%%%%%%
%%%%% Abstract %%%%%
%%%%%%%%%%%%%%%%%%%%
\begin{abstract}
The Fisher--Stefan model involves solving the Fisher--KPP equation on a domain whose boundary evolves according to a Stefan-like condition. The Fisher--Stefan model alleviates two practical limitations of the standard Fisher--KPP model when applied to biological invasion. First, unlike the Fisher--KPP equation, solutions to the Fisher--Stefan model have compact support, enabling one to define the interface between occupied and unoccupied regions unambiguously. Second, the Fisher--Stefan model admits solutions for which the population becomes extinct, which is not possible in the Fisher--KPP equation. Previous research showed that population survival or extinction in the Fisher--Stefan model depends on a critical length in one-dimensional Cartesian or radially-symmetric geometry. However, the survival and extinction behaviour for general two-dimensional regions remains unexplored. We combine analysis and level-set numerical simulations of the Fisher--Stefan model to investigate the survival--extinction conditions for rectangular-shaped initial conditions. We show that it is insufficient to generalise the critical length conditions to critical area in two-dimensions. Instead, knowledge of the region geometry is required to determine whether a population will survive or become extinct.
\end{abstract}
\end{frontmatter}
% \linenumbers
%%%%%%%%%%%%%%%%%%%%%%%%
%%%%% INTRODUCTION %%%%%
%%%%%%%%%%%%%%%%%%%%%%%%
\section{Introduction}\label{sec:intro}
The dimensionless Fisher--KPP equation,
\begin{equation}
  \label{eq:intro_fkpp}%
  \pd{u}{t} = \pdn{u}{x}{2} + u\left(1-u\right),
\end{equation}
is a classical prototype model in mathematical biology~\cite{Kolmogorov1937,Canosa1973,Grindrod1991,Murray2002} that describes the spatiotemporal evolution of a population density, \(u(x, t) > 0,\) that evolves due to linear Fickian diffusion combined with a logistic source term. A key property of the Fisher--KPP equation~\cref{eq:intro_fkpp} is that it admits travelling-wave solutions, with long-time speed \(c = 2\) for compactly-supported initial conditions, when solved on an infinite domain. By relating population invasion to travelling wave solutions, the Fisher--KPP equation and extensions of this model have been used to represent species invasion in ecology~\cite{Bradshaw-Hajek2004,Skellam1991,Shigesada1995}, front propagation in chemical reactions~\cite{Mercer1995}, and biological cell invasion~\cite{Gatenby1996,Maini2004,Sengers2007,Sherratt1990,Simpson2013,Treloar2014,Johnston2015}. However, despite its widespread usage, the Fisher--KPP equation has practical limitations. Firstly, any initial condition gives rise to population growth and complete colonisation. The Fisher--KPP equation is thus unsuitable for populations where extinction~\cite{El-Hachem2021} or arrested invasion~\cite{Landman2003} is of interest. A second shortcoming is that solutions to~\cref{eq:intro_fkpp} do not have compact support. Thus, we cannot identify the interface between occupied and unoccupied regions without ambiguity. This leads to difficulties in applying the Fisher--KPP equation to processes with well-defined invasion fronts, for example tumour cell invasion~\cite{Swanson2003,Perez-Garcia2011} or wound healing~\cite{Maini2004,McCue2019}.

To address the shortcoming of non-compactly-supported solutions, a common approach is to modify~\cref{eq:intro_fkpp} to incorporate degenerate nonlinear diffusion~\cite{Murray2002,Witelski1995}
\begin{equation}
  \label{eq:intro_nonlinear}%
  \pd{u}{t} = \pd{}{x}\left(D(u)\pd{u}{x}\right) + u\left(1-u\right),
\end{equation}
such that \(D(0) = 0.\) A common choice is \(D(u) = u^k\)~\cite{Gurney1975,Witelski1995} for \(k > 0.\) Setting \(k = 1\) leads to the \emph{Porous-Fisher's} equation~\cite{Witelski1995}, and for \(k > 1\)~\cref{eq:intro_nonlinear} is sometimes called the \emph{generalised Porous-Fisher's} equation~\cite{McCue2019,Warne2019}. Like the Fisher--KPP equation, the generalised Porous-Fisher's equation admits travelling-wave solutions. For the minimum wave speed, these travelling waves are sharp-fronted and have compact support~\cite{Sanchez-Garduno1994,Sherratt1996}, thus enabling one to define an unambiguous front. However, although~\cref{eq:intro_nonlinear} admits compactly-supported travelling-waves, like the Fisher--KPP equation it guarantees population survival, and cannot be used to model population extinction or receding fronts with local decrease in density~\cite{Landman2003,El-Hachem2021}. 

Another alternative to the standard Fisher--KPP equation is to consider a moving-boundary problem, such that~\cref{eq:intro_fkpp} holds on a compactly-supported region \(x < L(t),\) and \(u(L(t), t) = 0.\) The moving-boundary is assumed to evolve according to
\begin{equation}
  \label{eq:intro_stefan}
  \fd{L}{t} = -\kappa\pd{u}{x} \quad\text{ on }\quad x = L(t),
\end{equation}
where the parameter \(\kappa\) relates the density gradient at \(L(t)\) to the speed of the boundary. Moving-boundary problems of this type are traditionally used to model physical and industrial processes~\cite{Mitchell2009,Mitchell2010,Dalwadi2020,BrosaPlanella2019,BrosaPlanella2021}. Indeed, the boundary condition~\cref{eq:intro_stefan} is analogous to the classical Stefan condition~\cite{Crank1987} for a material undergoing phase change, where \(\kappa\) is the inverse Stefan number. More recently, moving-boundary problems have been used to study biological phenomena, including cancer invasion, cell motility, and wound healing~\cite{Shuttleworth2019,Kimpton2013,Fadai2020,Gaffney1999,Ward1997}. Applying the moving-boundary problem framework to~\cref{eq:intro_fkpp}, we obtain
\begin{subequations}
  \label{eq:intro_Fisher--Stefan}
  \begin{gather}
    \pd{u}{t} = \pdn{u}{x}{2} + u\left(1-u\right) \quad \text{ on } \quad 0 < x < L(t),\label{eq:intro_Fisher--Stefan_fkpp}\\
    \pd{u}{x} = 0 \quad \text{ on } \quad x = 0,  \label{eq:intro_Fisher--Stefan_no-flux}\\
    u = 0, \quad\fd{L}{t} = -\kappa\pd{u}{x} \quad \text{ on } \quad x = L(t), \label{eq:intro_Fisher--Stefan_interface}\\
    u(x,0) = u_0(x) \quad \text{ on } \quad 0 < x < L(0).\label{eq:intro_Fisher--Stefan_ic}
  \end{gather}
\end{subequations}
This extension to~\cref{eq:intro_fkpp}, known as the \emph{Fisher--Stefan} model, was first proposed by Du and Lin~\cite{Du2010}, and Du and Guo~\cite{Du2011,Du2012}. The Fisher--Stefan model alleviates the two practical disadvantages of Fisher--KPP, because the boundary \(L(t)\) defines the front position explicitly, and it admits solutions for population extinction~\cite{Du2010,Du2011,McCue2021}.

The Fisher--Stefan model has been studied extensively in one-dimensional Cartesian geometry. Du and Lin~\cite{Du2010} proved that~\cref{eq:intro_Fisher--Stefan} admits solutions whereby the population density evolves to a travelling wave solution as \(t \to \infty,\) with asymptotic speed that depends on \(\kappa.\) This corresponds to population survival and successful invasion, with \(u(x,t) \to 1\) as \(t \to\infty.\) However, Du and Lin~\cite{Du2010} also proved that the Fisher--Stefan model admits solutions where the population fails to establish, such that \(u \to 0^+\) as \(t \to\infty.\) This corresponds to population extinction, which cannot occur in the Fisher--KPP or generalised Porous-Fisher models. The survival--extinction behaviour of the Fisher--Stefan model depends on the critical length, \(L_c = \pi/2\)~\cite{Du2010,El-Hachem2019}. If \(L(0) > L_c,\) Du and Lin~\cite{Du2010} showed that the population will always survive. However, if \(L(0) < L_c,\) then survival only occurs if \(L(t)\) evolves such that \(L(t) > L_c\) at some time, and otherwise the population becomes extinct. In this \(L(0) < L_c\) scenario, whether the solution ever attains \(L(t) > L_c\) depends on the initial condition \(u_0(x),\) and the parameter \(\kappa\)~\cite{Du2010}.

Simpson~\cite{Simpson2020} extended the work of Du and Lin~\cite{Du2010} by considering the Fisher--Stefan model on an \(n\)-sphere. By replacing the second derivative term in~\cref{eq:intro_Fisher--Stefan_fkpp} with the \(n\)-dimensional radially-symmetric Laplacian operator, Simpson~\cite{Simpson2020} showed that a critical radius, \(R_c,\) governs survival and extinction analogously to the critical length. This critical radius depends on the dimension \(n.\) In the two-dimensional spreading-disc problem, the critical radius is \(R_c = \alpha_{01} \approx 2.4048,\) where \(\alpha_{01}\) is the first zero of \(J_0(x),\) the zeroth-order Bessel function of the first kind~\cite{Simpson2020}. However, the survival and extinction behaviour of the Fisher--Stefan model remains unexplored for non-radially-symmetric two-dimensional geometries. Investigating this is the subject of our study.

%%%%%%%%%%%%%%%%%%%%%%%%%%%%%%%%%%
%%%%% RESULTS AND DISCUSSION %%%%%
%%%%%%%%%%%%%%%%%%%%%%%%%%%%%%%%%%
\section{Results and Discussion}\label{sec:results}
\subsection{Mathematical Model}\label{ssec:results_model}
We consider a multidimensional extension to the dimensionless Fisher--Stefan model~\cref{eq:intro_Fisher--Stefan}. As illustrated in Figure~\cref{fig:results_numerics_level-set}A, we denote the region on which we solve the Fisher--KPP equation as \(\Omega(t),\) and its boundary as \(\uppartial\Omega(t).\) The boundary \(\uppartial\Omega(t)\) defines the interface between regions of non-zero population density and regions of zero density. Biologically, this might represent the position of a cell invasion front, or the boundary of a tumour. The multidimensional Fisher--Stefan model is then
\begin{subequations}
  \label{eq:results_Fisher--Stefan_2d}
  \begin{gather}
    \pd{u}{t} = \nabla^2u + u\left(1-u\right), \quad \text{ on } \quad \Omega(t),\label{eq:results_Fisher--Stefan_2d_fkpp}\\
    u = 0, \quad \text{ on } \quad \uppartial\Omega(t),\label{eq:results_Fisher--Stefan_2d_uf}\\
    V = -\kappa\nabla u \cdot\mathvec{\hat{n}} \quad \text{ on } \quad \uppartial\Omega(t),\label{eq:results_Fisher--Stefan_2d_stefan}\\
    u(\mathvec{x},0) = u_0(\mathvec{x}) \quad \text{ on } \quad \Omega(0),\label{eq:results_Fisher--Stefan_2d_ic}
  \end{gather}
\end{subequations}
where \(V\) is the normal speed of the interface, and \(\mathvec{\hat{n}}\) is the unit outward normal to the interface \(\uppartial\Omega(t).\) 
\subsection{Numerical Methods}\label{ssec:results_numerics}
\begin{figure}[htbp!]
  \centering
  \includegraphics[width=\linewidth]{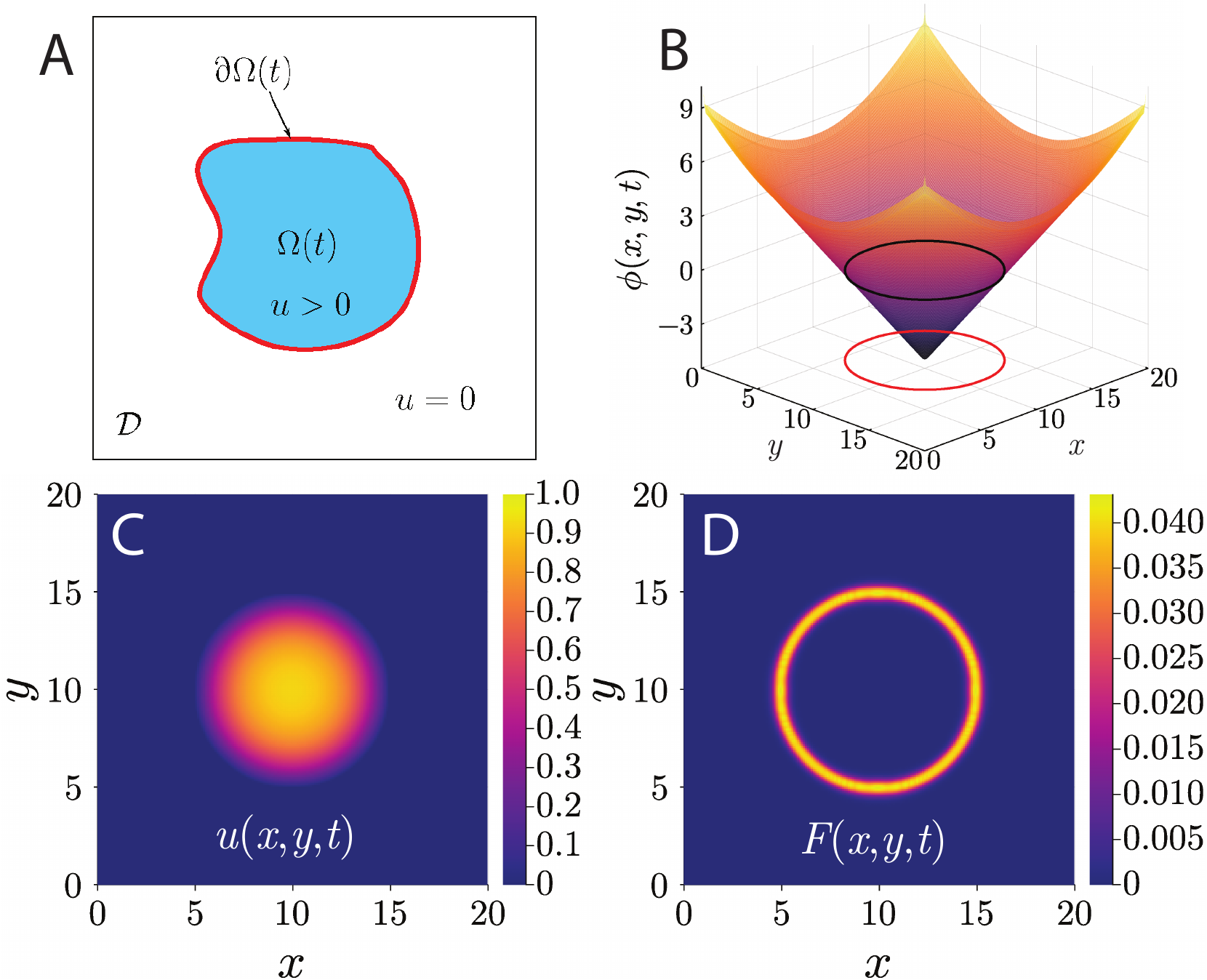}
  \caption{Illustration of the level-set method used to solve the two-dimensional Fisher--Stefan model~\cref{eq:results_Fisher--Stefan_2d}. (A) Schematic of the rectangular computational domain \(\mathcal{D},\) which contains the blue region \(\Omega(t)\) on which we solve the Fisher--KPP equation, and the red curve, which is the boundary \(\uppartial\Omega(t).\) (B) An example level-set function, \(\phi(x,y,t),\) for a disc-shaped \(\Omega(t).\) The black curve is the zero level-set, and the red curve is a projection onto the \((x,y)\) plane marked with the grid. (C) An example two-dimensional density profile \(u(x,y,t)\) for a circular region. (D) An example extension velocity field \(F(x,y,t).\)}
  \label{fig:results_numerics_level-set}
\end{figure}

We use the level-set method~\cite{Osher1988,Sethian1999,Osher2003,Aslam2004} to solve~\cref{eq:results_Fisher--Stefan_2d} on a two-dimensional computational domain, \(\mathcal{D}.\) This involves embedding the interface as the zero level-set of a scalar function \(\phi(\mathvec{x},t).\) That is,
\begin{equation}
  \label{eq:results_numerics_interface}%
  \uppartial\Omega(t) = \{\mathvec{x} \mid \phi\left(\mathvec{x},t\right) = 0\},
\end{equation}
where \(\phi(\mathvec{x},t)\) is defined for all \(\mathvec{x} \in \mathcal{D},\) and is such that \(\phi < 0\) for \(\mathvec{x} \in \Omega(t)\) and \(\phi \geq 0\) for \(\mathvec{x} \notin \Omega(t).\) An example for a disc is illustrated in Figure~\cref{fig:results_numerics_level-set}B. The level-set function \(\phi(\mathvec{x},t)\) evolves according to the level-set equation
\begin{equation}
  \label{eq:results_numericsl_level-set}%
  \pd{\phi}{t} + F\left\lvert\nabla\phi\right\rvert = 0,
\end{equation}
where \(F(\mathvec{x},t)\) is the \emph{extension velocity field}, a scalar function defined for \(\mathvec{x} \in \mathcal{D},\) such that \(F = V\) on \(\uppartial\Omega(t).\) An example extension velocity field for a spreading disc is shown in Figure~\cref{fig:results_numerics_level-set}D. As \(\phi\) evolves with time, so too does its zero level-set, which defines the region on which we solve~\cref{eq:results_Fisher--Stefan_2d_fkpp}. We solve the Fisher--Stefan model using explicit finite-difference methods~\cite{Tsitouras2011,Rackauckas2017,Lin2000,Jiang1998}.
For each numerical solution in this work, we use \(201\times 201\) spatial resolution, with \(\Delta t = 0.002,\) and \(\kappa = 0.1.\) This is sufficient to obtain good agreement with grid-independent spreading disc solutions of Simpson~\cite{Simpson2020}. Full details on our level-set method, an open-source \textsc{Julia} implementation, and numerical tests to validate the method are available on \href{https://github.com/alex-tam/2D_Fisher-Stefan_Level-Set}{Github}.
\subsection{Survival and Extinction in Two-Dimensions}\label{ssec:results_survival-extinction}
We begin our investigation of survival and extinction in two-dimensions by considering solutions with the disc initial condition,
\begin{equation}
  \label{eq:results_survival-extinction_ic_disc}%
  u(x,y,0) = \begin{cases} u_0 & \quad\text{ if }\quad \sqrt{\left(x - \frac{W}{2}\right)^2 + \left(y - \frac{W}{2}\right)^2} \leq R(0) \\ 0 & \quad\text{ otherwise,} \end{cases}
\end{equation}
where \(u_0\) is a constant satisfying \(0 < u_0 \leq 1,\) \(R(0)\) is the initial disc radius, and \(W\) is the width of the square computational domain \(\mathcal{D}.\) Simpson~\cite{Simpson2020} showed that the critical radius, \(R_c = \alpha_{01},\) explains the survival and extinction behaviour of a disc. With the initial condition~\cref{eq:results_survival-extinction_ic_disc}, the region \(\Omega(t)\) evolves as a disc of radius \(R(t).\) If \(R(t) > R_c\) at any time, the population survives. Alternatively, if \(R(t) < R_c\) for all \(t,\) then the population eventually becomes extinct. We solve the Fisher--Stefan model numerically using \(u_0 = 0.5,\) and compare solutions with \(R(0) = 2.1\) and \(R(0) = 2.4.\) Although, both values of \(R(0)\) are less than \(R_c,\) the Stefan-like condition~\cref{eq:results_Fisher--Stefan_2d_stefan} suggests that \(R(t)\) will increase for \(\kappa > 0,\) because \(\nabla u\cdot\hat{\mathvec{n}} < 0\) on \(\uppartial\Omega(t).\) However, it is unclear whether \(R(t)\) will ever exceed \(R_c,\) because the rate of spread \(V\) in~\cref{eq:results_Fisher--Stefan_2d_stefan} depends on \(\kappa\) and the shape of the density profile.

\begin{figure*}[htbp!]
  \centering
  \includegraphics[width=0.74\linewidth]{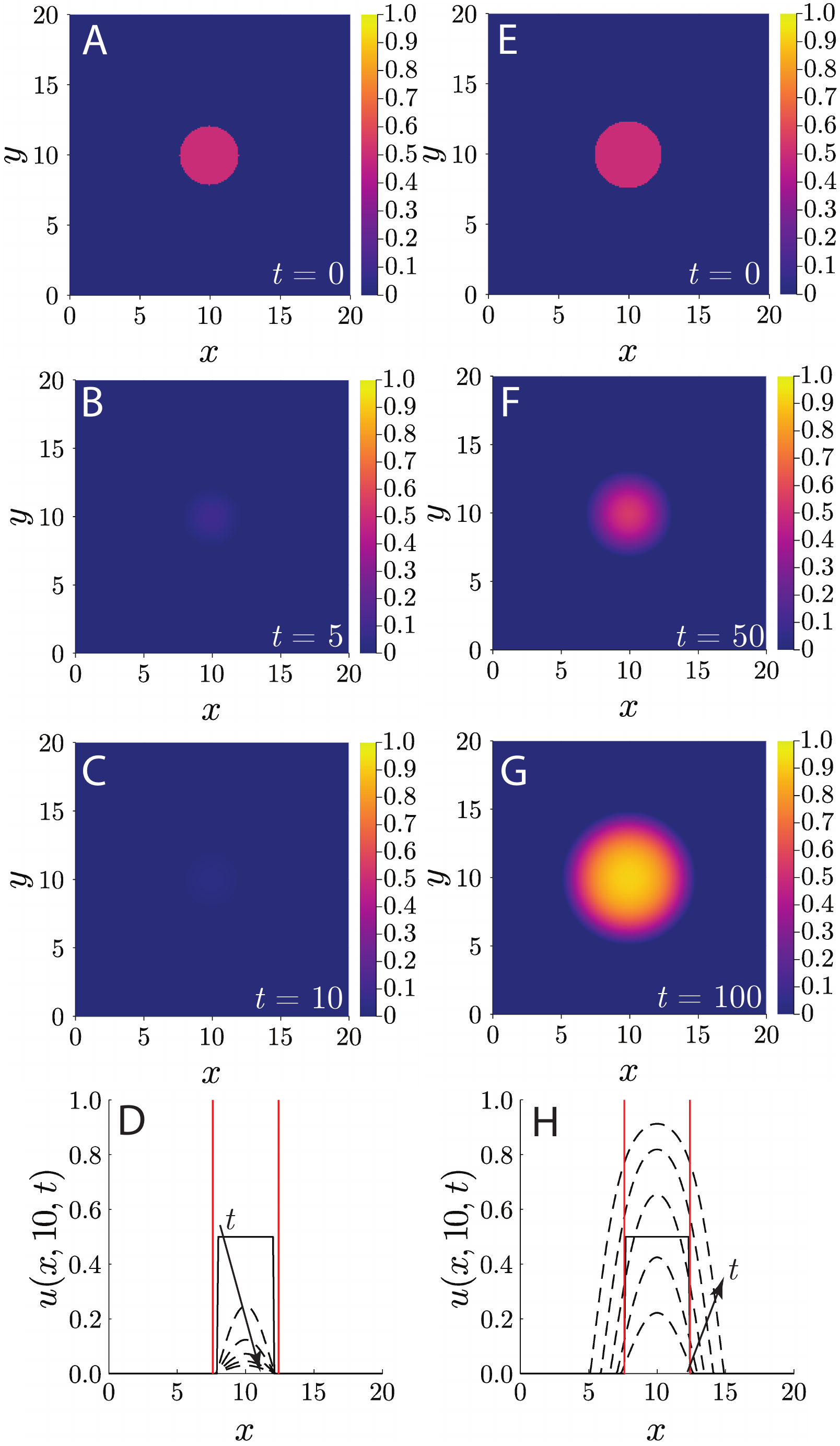}
  \caption{Numerical solutions to the Fisher--Stefan model with the disc initial condition~\cref{eq:results_survival-extinction_ic_disc}, and \(u_0 = 0.5.\) (A--D): Solution with \(R(0) = 2.1.\) (A): Initial condition, \(t = 0.\) (B): Density profile \(u(x,y,t)\) at \(t = 5.\) (C): Density profile \(u(x,y,t)\) at \(t = 10.\) (D): Evolution of population density, viewed as a one-dimensional slice through \(y = 10.\) The solid curve represents the initial condition, and dashed curves are the densities at \(t \in \{2,4,6,8,10\}.\) Arrow indicates direction of increasing \(t.\) (E--H): solution with \(R(0) = 2.4.\) (E): Initial condition, \(t = 0.\) (F): Density profile \(u(x,y,t)\) at \(t = 50.\) (G): Density profile \(u(x,y,t)\) at \(t = 100.\) (H): Evolution of population density, viewed as a one-dimensional slice through \(y = 10.\) The solid curve represents the initial condition, and dashed curves are the densities at \(t \in \{20,40,60,80,100\}.\) Arrow indicates direction of increasing \(t.\) (D\&H): Red vertical line represents \(x = 10 \pm R_c,\) where \(R_c = \alpha_{01}.\)}
  \label{fig:results_survival-extinction_disc}
\end{figure*}

Level-set numerical solutions with \(R(0) = 2.1\) and \(R(0) = 2.4\) are presented in Figure~\cref{fig:results_survival-extinction_disc}.
Panels A--D show that \(R(0) = 2.1\) results in population extinction, with density \(u \to 0^+\) as \(t\) increases. This is because \(R(t) < R_c,\) shown as coloured vertical lines in panel D. Eventually, population density \(u \to 0^+,\) which prevents further spread because \(\nabla u \to 0,\) and thus \(V \to 0\) according to~\cref{eq:results_Fisher--Stefan_2d_stefan}. In contrast, with \(R(0) = 2.4\) the population survives and spreads, such that both density \(u\) and the size of \(\Omega(t)\) eventually increase with time. Although the density \(u(10,10,t)\) at the centre of \(\Omega(t)\) decreases at first (see panel H), we eventually have \(R(t) > R_c.\) This enables the density to recover, and the population to survive. As \(t \to \infty,\) the solution approaches a radially-symmetric travelling wave. The ability to capture both survival and extinction is an advantage of the Fisher--Stefan model over the Fisher--KPP equation.

\clearpage

\begin{figure}[htbp!]
  \centering
  \includegraphics[width=0.6\linewidth]{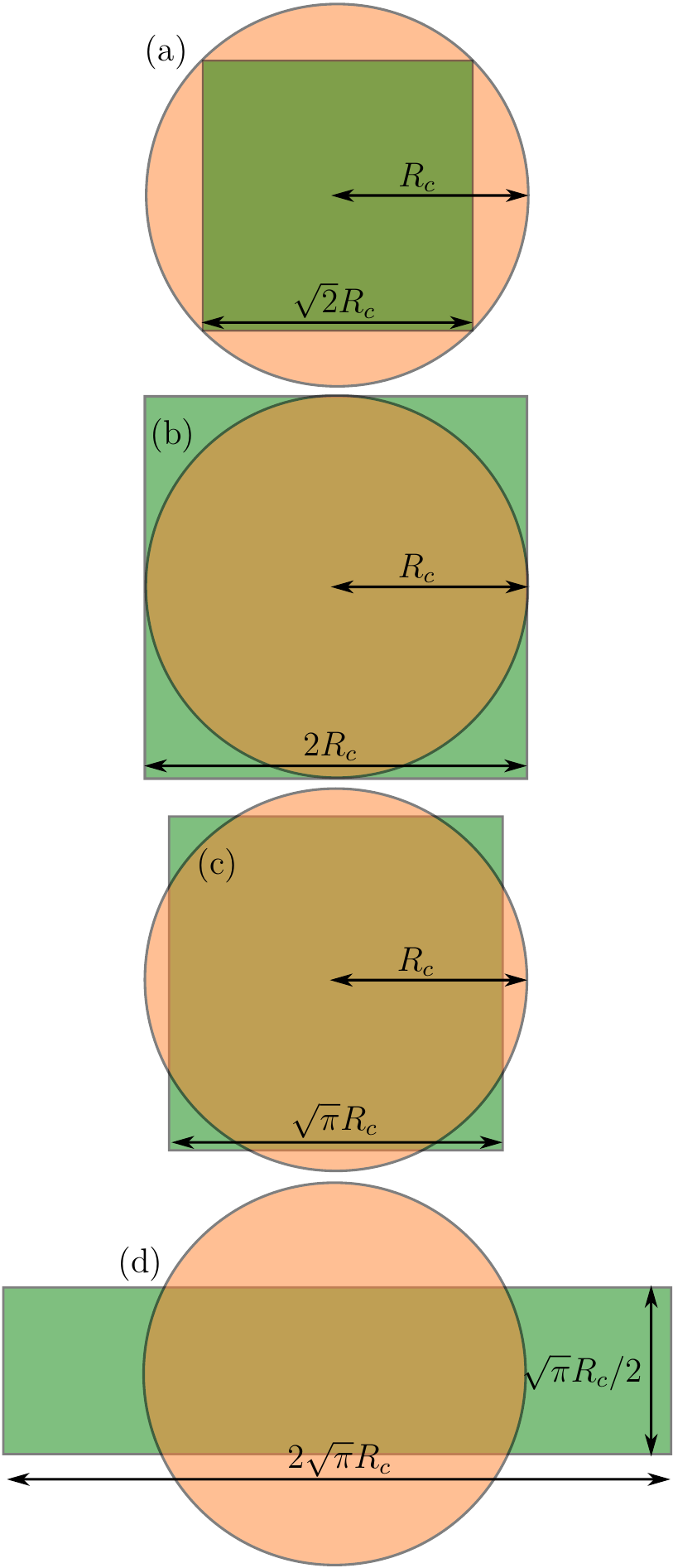}
  \caption{Potential square and rectangular analogues to the critical disc with radius \(R_c.\) (a) A square whose corners lie on the perimeter of the circle. (b) A square with side width equal to the circle diameter. (c) A square with the same area as the circle. (d) A rectangle with aspect ratio 4, and the same area as the circle.}
  \label{fig:results_survival-extinction_equivalent_rectangles}
\end{figure}

Despite being previously studied in radially-symmetric geometry, the survival and extinction behaviour of the Fisher--Stefan model in general two-dimensional geometry remains unexplored. For example, it is not immediately clear how the results for a disc might translate to rectangular shapes. Given a disc of radius \(R_c,\) one can conceive many rectangles that share certain properties with the disc. Some examples are shown in Figure~\cref{fig:results_survival-extinction_equivalent_rectangles}.
For example, in Figure~\cref{fig:results_survival-extinction_equivalent_rectangles}(a) we have the square with width \(\sqrt{2}R_c,\) for which all points inside the square are within the critical disc. In Figure~\cref{fig:results_survival-extinction_equivalent_rectangles}(b) we have the square with width \(2R_c,\) such that every point of the critical disc is contained within the square. Based on the radially-symmetric analysis, we would expect the initial square in Figure~\cref{fig:results_survival-extinction_equivalent_rectangles}(a) to give rise to extinction, and the initial square in Figure~\cref{fig:results_survival-extinction_equivalent_rectangles}(b) to give rise to survival. However, the situation is unclear for a square with side width between \(\sqrt{2}R_c\) and \(2R_c.\) One example is the square with width \(\sqrt{\pi}R_c\) (Figure~\cref{fig:results_survival-extinction_equivalent_rectangles}(c)), which has the same area as the critical disc. In this case, consider drawing a family of rays from the centre of the square or disc to the perimeter of each shape. Each ray is defined by its polar angle \(\theta \in [0, 2\pi].\) For some \(\theta,\) for example \(\theta = 0, \pm \pi/2,\) \(R_c\) exceeds the distance to the perimeter of the square. In contrast, for \(\theta = \pm\pi/4, \pm 3\pi/4,\) the distance to the perimeter of the square exceeds \(R_c.\) Therefore, it is unclear whether a population of this shape will survive or become extinct. Furthermore, varying the aspect ratio enables us to define a family of rectangles with the same area, one of which is shown in Figure~\cref{fig:results_survival-extinction_equivalent_rectangles}(d). The survival and extinction behaviour of these shapes is also unclear according to previous radially-symmetric analysis.

\begin{figure*}[htbp!]
  \centering
  \includegraphics[width=0.87\linewidth]{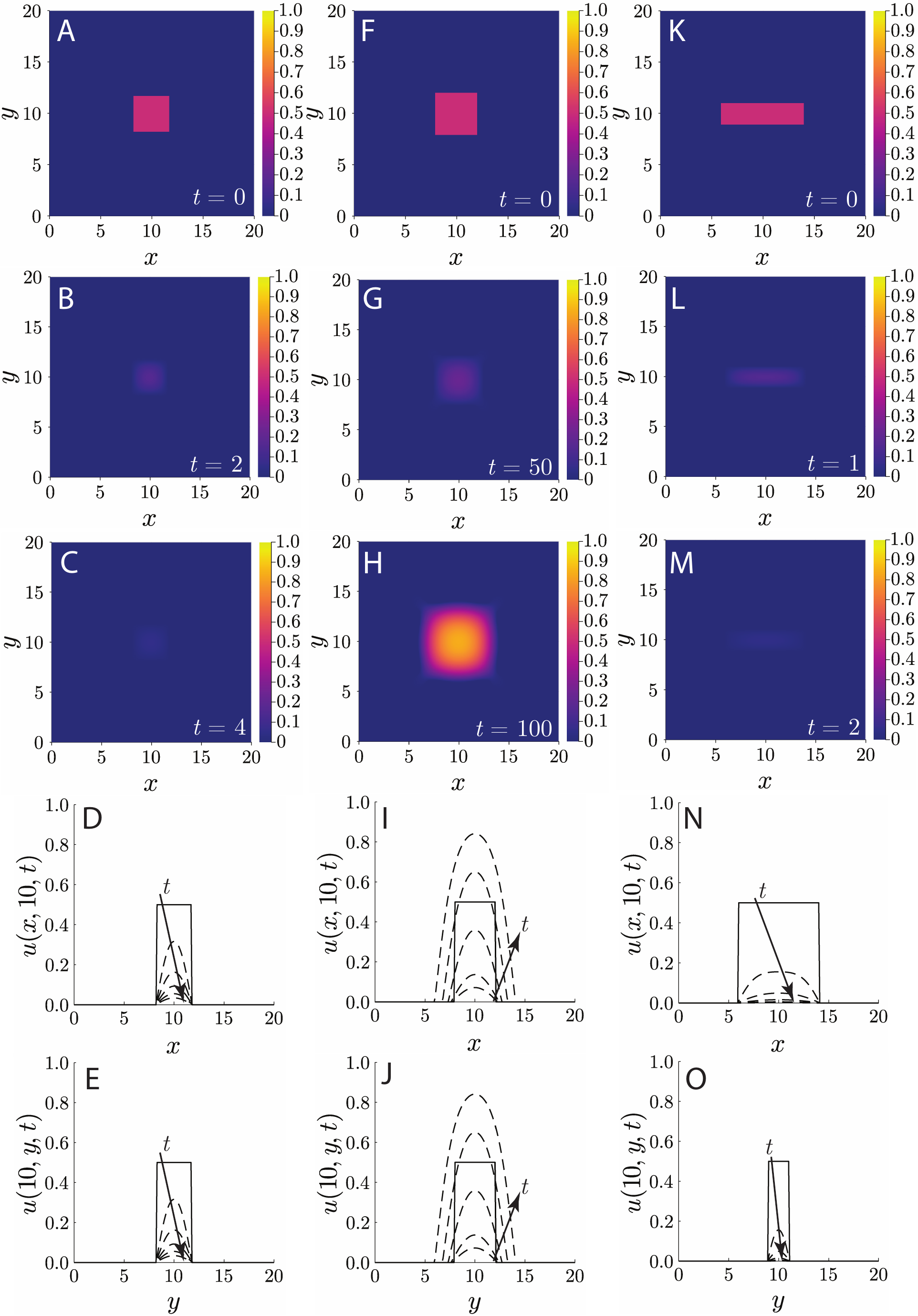}
  \caption{Numerical solutions with rectangular initial conditions~\cref{eq:results_survival-extinction_ic_rectangle}, with \(u_0 = 0.5.\) (A--E): Solution with \(L_x(0) = L_y(0) = 3.5.\) (F--J): Solution with \(L_x(0) = L_y(0) = 4.\) (K--O): Solution with \(L_x(0) = 8,\) \(L_y(0) = 2.\) (A): Initial density \(u(x,y,0).\) (B): Density profile at \(t = 2.\) (C): Density profile at \(t = 4.\) (D--E) One-dimensional slice \(u(x,10,t).\) \(u(10,y,t).\) Solid curve is the initial condition, dashed curves are solutions plotted at \(t \in \{1,2,3,4,5\}.\) Arrow indicates direction of increasing \(t.\) (F): Initial density \(u(x,y,0).\) (G): Density profile at \(t = 50.\) (H): Density profile at \(t = 100.\) (I--J) One-dimensional slice \(u(x,10,t).\) \(u(10,y,t).\) Solid curve is the initial condition, dashed curves are solutions plotted at \(t \in \{20,40,60,80,100\}.\) Arrow indicates direction of increasing \(t.\) (K): Initial density \(u(x,y,0).\) (L): Density profile at \(t = 1.\) (M): Density profile at \(t = 2.\) (N--O) One-dimensional slice \(u(x,10,t).\) \(u(10,y,t).\) Solid curve is the initial condition, dashed curves are solutions plotted at \(t \in \{1,2,3,4,5\}.\) Arrow indicates direction of increasing \(t.\)}
  \label{fig:results_survival-extinction_rectangles}
\end{figure*}

To address this question, we consider numerical solutions on initially rectangular domains. We characterise the geometry of \(\Omega(t)\) by defining \(L_x(t)\) and \(L_y(t)\) to be the distances in the \(x\) and \(y\)-directions respectively occupied by the population, measured through the centre of \(\Omega.\) That is,
\begin{subequations}
  \label{eq:results_survival_extinction_lengths}%
  \begin{align}
    L_x(t) &= \max\{x \mid u(x, W/2, t) > 0\} - \min\{x \mid u(x, W/2, t) > 0\},\label{eq:results_survival_extinction_lengths_Lx}\\
    L_y(t) &= \max\{y \mid u(W/2, y, t) > 0\} - \min\{y \mid u(W/2, y, t) > 0\}.\label{eq:results_survival_extinction_lengths_Ly}
  \end{align}
\end{subequations}
For initially-rectangular domains, we then have the initial condition
\begin{equation}
  \label{eq:results_survival-extinction_ic_rectangle}%
  u(x,y,0) = \begin{cases} u_0 & \quad\text{ if }\quad \left\lvert x - \frac{W}{2}\right\rvert \leq \frac{L_x(0)}{2}\quad\text{ and }\quad \left\lvert y - \frac{W}{2}\right\rvert \leq \frac{L_y(0)}{2},  \\ 0 & \quad\text{ otherwise.} \end{cases}
\end{equation} 
Numerical results are shown in Figure~\cref{fig:results_survival-extinction_rectangles}. Panels A--E show that a solution with the square initial condition \(L_x(0) = L_y(0) = 3.5\) leads to extinction. Increasing the size of the initial square to \(L_x(0) = L_y(0) = 4\) yields survival (Figure~\cref{fig:results_survival-extinction_rectangles}, panels F--J). A rectangular initial condition with \(L_x(0) = 8\) and \(L_y(0) = 2\) leads to extinction, as panels K--O of Figure~\cref{fig:results_survival-extinction_rectangles} show. This occurs despite this rectangle having the same initial area as the square that survived. Thus, the survival or extinction of initially-rectangular regions does not depend solely on area, and more detail about the region geometry is required.

To provide insight into the results in Figure~\cref{fig:results_survival-extinction_rectangles}, we perform similar analysis to El-Hachem \emph{et al.}~\cite{El-Hachem2019} and Simpson~\cite{Simpson2020}. This involves considering the limit \(u \to 0^+,\) and obtaining critical conditions that determine survival or extinction. In this limit, the leading-order population density \(\hat{u}(x,y,t)\) satisfies
\begin{equation}
  \label{eq:results_survival-extinction_fkpp_linearised}%
  \pd{\hat{u}}{t} = \pdn{\hat{u}}{x}{2} + \pdn{\hat{u}}{y}{2} + \hat{u}.
\end{equation}
We consider solutions to~\cref{eq:results_survival-extinction_fkpp_linearised} on a fixed rectangular domain \(0 \leq x \leq X\) and \(0 \leq y \leq Y,\) subject to arbitrary compactly-supported initial conditions, and homogeneous Dirichlet boundary conditions on all boundaries. This problem has the general solution
\begin{equation}
  \label{eq:results_survival-extinction_gs}%
    \hat{u}(x,y,t) = \sum_{n=1}^{\infty}\sum_{m=1}^{\infty} A_{n,m}\sin\left(\frac{n\pi x}{X}\right)\sin\left(\frac{m\pi y}{Y}\right)\e^{-\left(\frac{n^2\pi^2}{X^2} + \frac{m^2\pi^2}{Y^2} - 1\right)t},
\end{equation}
with coefficients \(A_{n,m}\) chosen such that \(\hat{u}(x,y,0)\) matches the initial condition. We next consider the long-time solution \(t \to \infty,\) such that the leading-eigenvalue \(n = m = 1\) approximates the general solution~\cref{eq:results_survival-extinction_gs}. Then,
\begin{equation}
  \label{eq:results_survival-extinction_leading_eigenvalue}%
  \hat{u}(x,y,t) \thicksim A_{1,1}\sin\left(\frac{\pi x}{X}\right)\sin\left(\frac{\pi y}{Y}\right)\e^{-\left(\frac{\pi^2}{X^2} + \frac{\pi^2}{Y^2} - 1\right)t}\quad\text{ as } \quad t \to\infty.
\end{equation}
To obtain the conditions under which the population survives or becomes extinct as \(t \to\infty,\) we use the approximation~\cref{eq:results_survival-extinction_leading_eigenvalue} and consider conservation of the population inside \(\Omega.\) For \(u \ll 1,\) the conservation statement is
\begin{equation}
  \label{eq:results_survival-extinction_conservation}%
  \fd{M}{t} = \int_{\Omega} \hat{u}(x,y,t) - \int_{\uppartial\Omega} -\nabla \hat{u}\cdot\mathvec{\hat{n}},
\end{equation}
where \(M(t) = \int_{\Omega} \hat{u}(x,y,t)\) is the total population in \(\Omega(t).\) The first term on the right-hand side of~\cref{eq:results_survival-extinction_conservation} is the net accumulation of \(\hat{u}\) inside \(\Omega(t)\) due to the (linearised) source term. The second term on the right-hand side of~\cref{eq:results_survival-extinction_conservation} is the rate of population loss through the boundary \(\uppartial\Omega(t)\) due to diffusion. The population survives if \(\d M/\d t > 0\) as \(t \to\infty,\) that is  accumulation exceeds loss. Alternatively, the population becomes extinct if \(\d M/\d t < 0\) as \(t \to\infty,\) \emph{i.e.} the rate of loss exceeds accumulation. Critical behaviour occurs when \(\d M/\d t = 0\) as \(t \to\infty,\) such that the rates of population gain and loss balance. This requires
\begin{equation}
  \label{eq:results_survival-extinction_conservation_critical}%
  \int_{\Omega} \hat{u}(x,y,t) = \int_{\uppartial\Omega} -\nabla \hat{u}\cdot\mathvec{\hat{n}}.
\end{equation}
In fixed rectangular geometry, \(\Omega\) and \(\uppartial\Omega\) are readily parameterised to give
\begin{equation}
  \label{eq:results_survival-extinction_conservation_critical_expanded}%
  \begin{gathered}
    \int_{0}^{Y} \int_{0}^{X} \hat{u}(x,y,t) \df{x}\df{y} = -\int_{0}^{Y} \pd{\hat{u}(X,y,t)}{x} \df{y} \\
    + \int_{0}^{Y} \pd{\hat{u}(0,y,t)}{x} \df{y} - \int_{0}^{X} \pd{\hat{u}(x,Y,t)}{y}\df{x} + \int_{0}^{X} \pd{\hat{u}(x,0,t)}{y} \df{x}.
  \end{gathered}
\end{equation}
Substituting the leading-eigenvalue approximation~\cref{eq:results_survival-extinction_leading_eigenvalue} for \(\hat{u}\) in~\cref{eq:results_survival-extinction_conservation_critical_expanded}, evaluating integrals, and simplifying then yields
\begin{equation}
  \label{eq:results_survival-extinction_critical_condition}%
  XY = \pi\sqrt{Y^2 + X^2}.
\end{equation}
The condition~\cref{eq:results_survival-extinction_critical_condition} implies that the critical area for a rectangle with dimensions \(L_x \times L_y\) is \(A_c = \pi\sqrt{L_x^2 + L_y^2}.\) The analysis suggests that a rectangular population will survive if its lengths \(L_x(t)\) and \(L_y(t)\) ever satisfy \(L_xL_y > A_c,\) and will become extinct if  \(L_xL_y < A_c\) for all time. Another interpretation of the condition~\cref{eq:results_survival-extinction_critical_condition} is that extinction will occur if the exponential term in~\cref{eq:results_survival-extinction_leading_eigenvalue} decays, and the population will survive if the exponential term grows. Setting the exponent in the leading-eigenvalue approximation~\cref{eq:results_survival-extinction_leading_eigenvalue} to zero then yields the same condition~\cref{eq:results_survival-extinction_critical_condition}.

Based on our analysis, the inequalities
\begin{equation}
  \label{eq:results_survival-extinction_critical_inequality}%
  L_y > \pi\sqrt{\frac{L_x^2}{L_x^2-\pi^2}}, \quad L_x > \pi,
\end{equation}
must be satisfied at some \(t\) for population survival. Interestingly, the population cannot survive if either \(L_x\) or \(L_y\) remain less than \(\pi\) for all time. Furthermore, for each \(L_x > \pi,\) there exists a unique critical \(L_{y,c}\) given by the right-hand side of the first inequality in~\cref{eq:results_survival-extinction_critical_inequality}, above which the population will survive. This completely characterises the effect of rectangular geometry and aspect ratio on survival and extinction in the two-dimensional Fisher--Stefan model. The conditions~\cref{eq:results_survival-extinction_critical_inequality} enable comparison with the radially-symmetric results of Simpson~\cite{Simpson2020}. The critical disc with radius \(R_c = 2.4048\) has area \(A = 18.17.\) By comparison, the critical square has widths \(L_x = L_y = \sqrt{2}\pi \approx 4.443,\) and area \(A = 19.74.\)

\begin{figure}[htbp!]
  \centering
  \includegraphics[width=\linewidth]{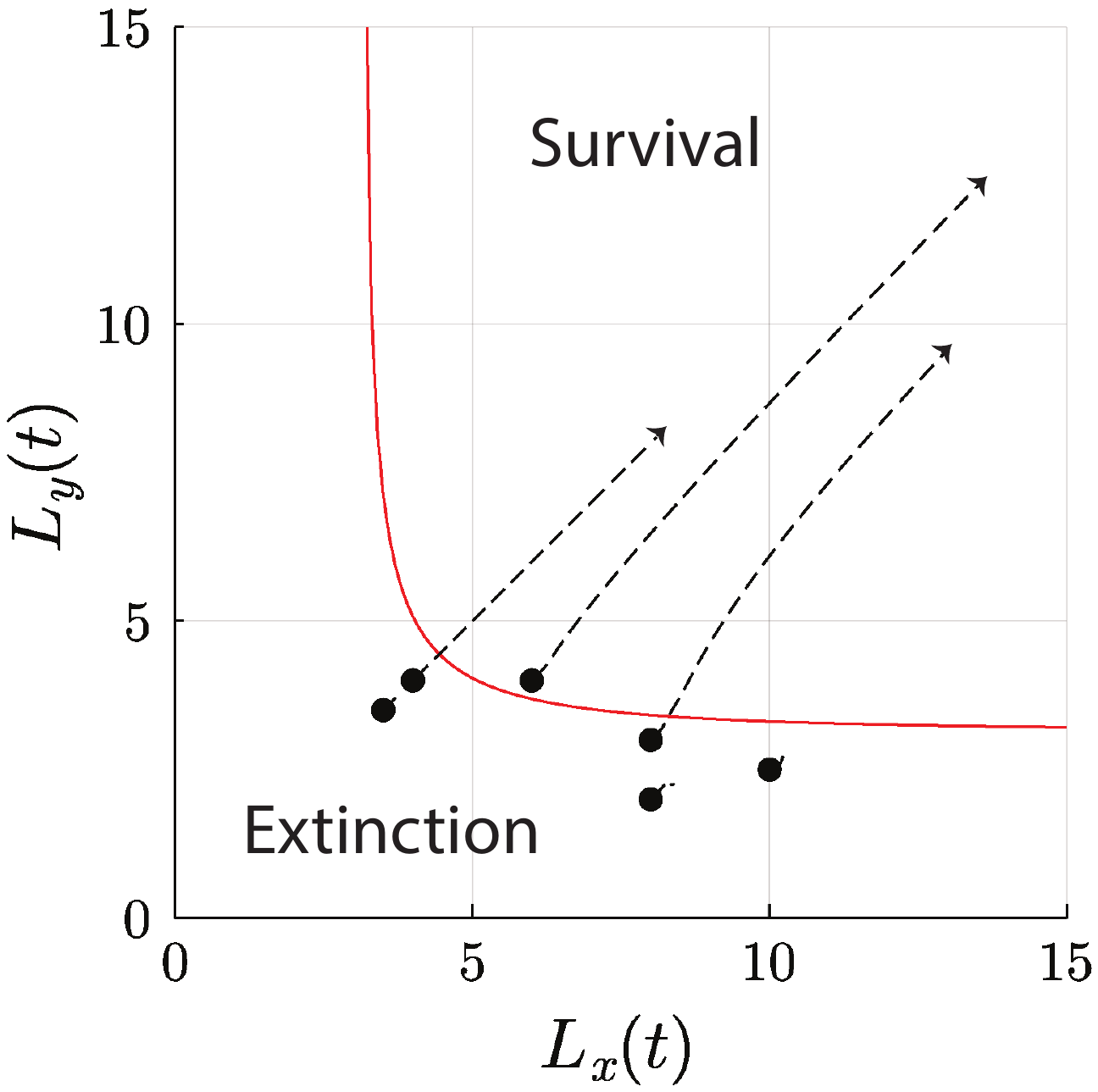}
  \caption{Evolution of numerical solutions to the Fisher--Stefan model with rectangular initial conditions. Black dots represent the initial rectangle dimensions \(L_x(0),\) and \(L_y(0)\) for each solution plotted. Dashed curves represent the evolution of \(L_x(t)\) and \(L_y(t)\) from \(t = 0\) to \(t = 10\) for extinction solutions, or \(t = 0\) to \(t = 100\) for survival solutions. The red curve defines the critical threshold between survival and extinction, as per~\cref{eq:results_survival-extinction_critical_inequality}.}
  \label{fig:results_survival-extinction_analysis_comparison}
\end{figure}

Our analysis explains the numerical solutions in Figure~\cref{fig:results_survival-extinction_rectangles}. For the square solution in Figure~\cref{fig:results_survival-extinction_rectangles}, panels A--E with \(L_x(0) = L_y(0) = 3.5,\) the widths never reach \(\sqrt{2}\pi,\) and thus the population eventually becomes extinct. However, with \(L_x(0) = L_y(0) = 4,\) (Figure~\cref{fig:results_survival-extinction_rectangles}, panels F--J), the square widths eventually exceed \(\sqrt{2}\pi,\) leading to eventual survival. For a rectangle with \(L_x(0) = 8\) and \(L_y(0) = 2,\) the population becomes extinct because \(L_y(t) < \pi\) for all \(t,\) which guarantees extinction. These and other results are summarised in Figure~\cref{fig:results_survival-extinction_analysis_comparison}, which illustrates how numerical solutions with rectangular initial conditions evolve. If the inequalities~\cref{eq:results_survival-extinction_critical_inequality} holds for any \(t,\) the population survives and \(L_x\) and \(L_y\) continue to increase. This is shown by the trajectories above the red curve in Figure~\cref{fig:results_survival-extinction_analysis_comparison}. Conversely, the population becomes extinct if~\cref{eq:results_survival-extinction_critical_inequality} never holds. This occurs for the trajectories that remain below the red curve in Figure~\cref{fig:results_survival-extinction_analysis_comparison}.
The critical conditions~\cref{eq:results_survival-extinction_critical_inequality} hold provided the domain \(\Omega(t)\) remains approximately rectangular. Our choice of \(\kappa = 0.1\) in all numerical solutions of Figures~\cref{fig:results_survival-extinction_rectangles,fig:results_survival-extinction_analysis_comparison} ensures that the rectangular shapes persist sufficiently long for~\cref{eq:results_survival-extinction_critical_inequality} to determine survival or extinction. With larger \(\kappa,\) for example \(\kappa = 1,\) initially-rectangular shapes satisfying the survival conditions rapidly evolve to an expanding disc. In these scenarios, the long-term survival--extinction behaviour instead obeys the critical radius condition of Simpson~\cite{Simpson2020}.
\section{Conclusion and Future Work}\label{sec:conclusion}
The Fisher--Stefan model provides an alternative to standard the Fisher--KPP model that both defines an unambiguous invasion front, and admits solutions whereby the population becomes extinct. In this study, we extended one-dimensional and radially-symmetric results for the Fisher--Stefan model to two-dimensional rectangular domains. Using the level-set method, we computed numerical solutions to the two-dimensional Fisher--Stefan model with \(\kappa = 0.1,\) for radially-symmetric and rectangular domains. Our objective was to understand the conditions under which square and rectangular shapes give rise to survival or extinction.

To explain the numerical observations, we obtained conditions for the Fisher--Stefan model that govern survival and extinction of a rectangular-shaped population. Using a conservation argument in the small density limit, we showed that survival requires both lengths of the rectangle to exceed \(\pi,\) and for the rectangle area to exceed a critical threshold that depends on the side lengths. Extending the one-dimensional idea of critical length or critical radius to a critical area in two dimensions is insufficient to explain survival and extinction. Instead, for rectangular-shaped domains information about both widths is necessary. Interestingly, these findings are similar to the behaviour of a population subject to the strong Allee effect. In radially-symmetric geometry, Lewis and Kareiva~\cite{Lewis1993} showed that a critical radius governs survival and extinction. However, recent work by Li \emph{et al.}~\cite{Li2021} showed that additional geometric information is required to determine survival or extinction in asymmetric two-dimensional rectangular domains.

The analytical methods established in this work apply directly to any domain \(\Omega(0)\) such that solution to~\cref{eq:results_survival-extinction_fkpp_linearised} with homogeneous Dirichlet conditions on \(\uppartial\Omega(0)\) can be obtained using separation of variables. One avenue for future work is to investigate the conditions for survival and extinction in these separable shapes, for example an ellipse~\cite{Abramowitz1964}. Furthermore, our numerical method can be used to study survival and extinction for general initial shapes \(\Omega(0),\) regardless of whether these are separable. This includes investigation of non-simply-connected initial shapes.

All numerical solutions presented in this work have \(\kappa = 0.1.\) When solving the Fisher--Stefan model numerically with larger \(\kappa,\) for example \(\kappa = 1,\) we observed that initially square and rectangular domains that satisfied the conditions for survival quickly evolved to an expanding circular shape. This numerical evidence suggests that radially-symmetric survival solutions to the Fisher--Stefan model are stable with respect to small-amplitude azimuthal perturbations. However, this has not been verified analytically. The stability of inward-moving fronts in a hole-closing geometry~\cite{McCue2019} is also yet to be analysed. We plan to investigate this stability problem using analytical and numerical methods in future work.
%
%%%%%%%%%%%%%%%%%%%%%%%
%%%%% Back Matter %%%%%
%%%%%%%%%%%%%%%%%%%%%%%
\section*{Acknowledgements}
Computational resources and services used in this work were provided by HPC and Research Support Group, Queensland University of Technology, Brisbane, Australia. M. J. S. also acknowledges funding from the Australian Research Council (Grant No. DP200100177).
\section*{Author Contributions}
M. J. S. designed the research. A. K. Y. T. wrote the numerical code, obtained the results, performed the analysis, and wrote the manuscript, with guidance from M. J. S.
\section*{Competing Interests}
We have no competing interests to declare.
%%%%%%%%%%%%%%%%%%%%%%%%
%%%%% BIBLIOGRAPHY %%%%%
%%%%%%%%%%%%%%%%%%%%%%%%
\bibliography{2d_survival_extinction}
\end{document}